# Fused Analytical and Iterative Reconstruction (AIR) via modified proximal forward-backward splitting: a FDK-based iterative image reconstruction example for CBCT

Hao Gao[1]

[1]Department of Radiation Oncology, Duke University Medical Center, Durham, NC 27710, USA

E-mail: hao.gao.2012@gmail.com

**Abstract.** This work is to develop a general framework, namely analytical iterative reconstruction (AIR) method, to incorporate analytical reconstruction (AR) method into iterative reconstruction (IR) method, for enhanced CT image quality and reconstruction efficiency. Specifically, AIR is established based on the modified proximal forward-backward splitting (PFBS) algorithm, and its connection to the filtered data fidelity with sparsity regularization is discussed.

As a result, AIR decouples data fidelity and image regularization with a two-step iterative scheme, during which an AR-projection step updates the filtered data fidelity term, while a denoising solver updates the sparsity regularization term. During the AR-projection step, the image is projected to the data domain to form the data residual, and then reconstructed by certain AR to a residual image which is then weighted together with previous image iterate to form next image iterate. Intuitively since the eigenvalues of AR-projection operator are close to the unity, PFBS based AIR has a fast convergence. Such an advantage is rigorously established through convergence analysis and numerical computation of convergence rate.

The proposed AIR method is validated in the setting of circular cone-beam CT with AR being FDK and total-variation sparsity regularization, and has improved image quality from both AR and IR. For example, AIR has improved visual assessment and quantitative measurement in terms of both contrast and resolution, and reduced axial and half-fan artifacts.

## 1. Introduction

In the seminal work on simultaneous algebraic reconstruction technique (SART) [1], a trick for improving reconstructed image quality that might be easily overlooked is through the use of filtering to weight backprojection by a Hamming window, which is a kind of backprojection-filtration process [2,3]. In the so-called iterative filtered backprojection (FBP) (IFBP) that was originally developed for attenuation correction in SPECT [4,5], a FBP-projection or backprojection-filtering-projection iterative scheme was proposed with improved reconstructed image quality. For improved CT reconstruction, IFBP or its variants were developed for limited data problem [6], parallel-beam geometry [7], cone-beam CT (CBCT) [8] with one iteration based on FDK, helical CT [9] using a weighted FBP method. In addition, IFBP converges fast since its eigenvalues are clustered together around the unity [10,11]. All of these suggest that the incorporation of filtering into iterative method can improve both reconstruction accuracy and speed.

In recent years, there have been tremendous developments in CT image reconstruction algorithms in terms of both analytical reconstruction (AR) methods and iterative reconstruction (IR) methods, particularly sparsity-regularized model-based IR methods [12,13] inspired by compressive sensing [14,15] for a wide range of CT problems [16-23]. For the purpose of synergizing AR and IR, we investigated a filtration-weighted formulation of data fidelity with sparsity regularization in the setting of 2D fan-beam CT and developed the image reconstruction algorithm based on alternating direction method of multipliers (ADMM) [24] or split Bregman method [25], i.e., so-called fused analytical and iterative reconstruction (AIR) [26,27]. Note that the iterative step derived from the L2 subproblems of ADMM for AIR is similar to the IFBP iteration, although the backprojection in the AIR scheme does not have to include projection pre-weighting, modified filter for fan-beam geometry, or angular weighting in backprojection as in FBP or IFBP since the filtering operator is the crucial component.

This work develops AIR using a modified framework of proximal forward-backward splitting (PFBS) [28], which completely decouples data fidelity minimization and image regularization problems. A 3D example of FDK-based AIR will be developed for CBCT. The convergence of PFBS-based AIR will be rigorously established that accommodates the incorporation of general AR methods into IR. On the other hand, the connection of AIR to the filtered data fidelity will be discussed in the sense of unmatched projection/backprojection pairs [10,11].

As a result, a backprojection-filtering-projection or in general AR-projection step with no regularization operators that enjoys the fast convergence similar to IFBP can be derived for data fidelity minimization, and then sole image regularization or denoising subproblems with no projection operators can be rapidly solved by ADMM. This is considerably faster than entirely ADMM-based optimization algorithm without filtering, owning to the fast convergence from the incorporation of filtering. In addition, based on PFBS, the proposed filtering based iterative method can be readily connected to existing iterative methods [10] and serve as a framework for conveniently synergizing AR and IR.

## 2. Methods

### 2.1. Proximal Forward-Backward Splitting

In this section, we review the IR method based on proximal forward-backward splitting (PFBS) to motivate the AIR method via PFBS.

The IR method under consideration is formulated as the following optimization problem

$$x^* = \arg\min_{x} f(x) + g(x). \tag{1}$$

To simplify the discussion, here $f(x)$ is a least-square data fidelity and $g(x)$ is the image regularization term via isotropic total variation (TV) that is commonly used in IR methods for CT [12,13], i.e.,

$$f(x) = \frac{1}{2} \| Ax - y \|_2^2 \tag{2}$$

and

$$g(x) = \lambda \| \nabla x \|_1. \tag{3}$$

In the above, $x$ is the image to be reconstructed, $y$ the projection data, $A$ the system matrix or a projection operator discretized from the X-ray transform, $\lambda$ is a regularization parameter or the model parameter to balance data fidelity and image regularization, $\|\cdot\|_2$ the L2 norm, the discrete TV transform

$$[\nabla x]_{ijk} = (z_{1,ijk}, z_{2,ijk}, z_{3,ijk}) = (x_{i+1,jk} - x_{ijk}, x_{i,j+1,k} - x_{ijk}, x_{ij,k+1} - x_{ijk}) \tag{4}$$

and isotropic TV norm

$$\| \nabla x \|_1 = \sum_{ijk} \sqrt{z_{1,ijk}^2 + z_{2,ijk}^2 + z_{3,ijk}^2}, \tag{5}$$

where $i$, $j$, and $k$ are the voxel indexes of $x$ along three Cartesian coordinates respectively.

Next we review the PFBS method for solving Eq. (1) by first describing the proximal operator, which was introduced by Moreau [44], i.e.,

$$prox_g(y) = \arg\min_x \frac{1}{2} \| x - y \|_2^2 + g(x). \tag{6}$$

Here $g$ is a closed proper convex function (e.g., the TV norm Eq. (5)). And the minimizer of Eq. (6) denoted by $prox_g$ for the given data $y$ is unique and often formally denoted by

$$prox_g(y) = (1 + \partial g)^{-1} y, \tag{7}$$

which can be derived from the optimal condition of Eq. (6). For non-differentiable functions (e.g., the TV norm Eq. (5)), $\partial g$ In Eq. (7) can be understood in the sense of a subdifferential operator.

Then the PFBS algorithm or so-called proximal gradient method [28] for minimizing Eq. (1) mainly consists of two iterative steps with $x^0=0$

$$x^{n+1/2} = x^n - s\nabla f(x^n), \tag{8}$$

$$x^{n+1} = prox_{sg}(x^{n+1/2}). \tag{9}$$

A nice feature of the PFBS algorithm is that it completely decouples the IR into two separate subproblems: data fidelity minimization Eq. (8) and image denoising Eq. (9). This decoupling has also been considered previously in CT [12,13], and the convergence of PFBS can be rigorously justified [28].

For a self-contained explanation, the above algorithm will be formally derived from the point view of a fixed-point iteration [45]. That is, $x^*$ is a solution of Eq. (1) if and only if in the sense of subdifferential operators

$$0 \in \nabla f(x^*) + \partial g(x^*), \tag{10}$$

which can be rewritten as

$$x^* - s\nabla f(x^*) \in x^* + s\partial g(x^*), \tag{11}$$

where s is an algorithm parameter. Thus, from Eq. (11) and the uniqueness of the proximal operator Eq. (7), the optimizer of Eq. (1) can be characterized as a fixed point of the following forward-backward operator

$$x^* = (1 + s\partial g)^{-1}(x^* - s\nabla f(x^*)). \tag{12}$$

Therefore, we arrive at the iterative scheme Eq. (8) and Eq. (9) to solve for $x^*$, i.e.,

$$x^{n+1/2} = x^n - s^n A^T(Ax^n - y), \tag{13}$$

$$x^{n+1} = \arg\min_x \frac{1}{2} \| x - x^{n+1/2} \|_2^2 + s^n \lambda \| \nabla x \|_1. \tag{14}$$

*2.2. Modified Proximal Forward-Backward Splitting*

In this section we introduce a variant of the PFBS method Eq. (13) and Eq. (14), namely, the modified PFBS method, based on which the AIR method will be developed in the next section. In comparison with the PFBS method Eq. (13) and Eq. (14), the modified PFBS method consists of the following iterative steps with $x^0=0$

$$x^{n+1/2} = x^n - s^n F(Ax^n - y), \tag{15}$$

$$x^{n+1} = \arg\min_x \frac{1}{2} \| x - x^{n+1/2} \|_2^2 + s^n \lambda \| \nabla x \|_1. \tag{16}$$

Note that the only difference between PFBS and the modified PFBS is that $A^T$ in Eq. (13) is replaced by $F$ in Eq. (15).

It is worthy to note that the modified PFBS does not have a corresponding optimization problem to solve for, while PFBS solves a corresponding optimization problem Eq. (1). For example, in the practical PFBS implementation, $A^T$ may not be strictly the transpose of $A$, e.g., $A$ is discretized from the ray-driven method while $A^T$ is discretized from the pixel-driven method to avoid the numerical artifacts [31]. Therefore strictly speaking, the practical PFBS implementation in this case is also a modified PFBS method, and it does not have a corresponding optimization problem either.

Despite the non-existence of a corresponding optimization problem, the modified PFBS method Eq. (15) and Eq. (16) is still convergent. This is intuitively true for the practical PFBS implementation in which $A^T$ is not strictly the transpose of $A$. The rigorous convergence analysis is provided next regarding the general situation.

Let us rewrite Eq. (15) and Eq. (16) using the proximal operator Eq. (7), i.e.,

$$x^{n+1} = prox_{sg}(x^n - sF(Ax^n - y)), \tag{17}$$

where $g$ is the isotropic TV norm Eq. (3). Here for simplicity, $s$ is assumed to be a constant, which also holds in implementation. Then we have

$$\begin{aligned} \| x^{n+1} - x^n \| &= \| prox_{sg}(x^n - sF(Ax^n - y)) - prox_{sg}(x^{n-1} - sF(Ax^{n-1} - y)) \| \\ &\leq \| (I - sFA)(x^n - x^{n-1}) \| \\ &\leq \| I - sFA \| \cdot \| x^n - x^{n-1} \| \end{aligned}, \tag{18}$$

where the first inequality is from the non-expansion property of $prox_{sg}$. Thus it is clear then the sequence $\{x^n\}$ is convergent if

$$\| I - sFA \| < 1. \tag{19}$$

Moreover, let $x^*$ be the convergent point of $\{x^n\}$, by the similar argument for Eq. (18) we have

$$\begin{aligned} \| x^{n+1} - x^* \| &\leq \| (I - sFA)(x^n - x^*) \| \\ &\leq \| I - sFA \|^{n+1} \cdot \| x^0 - x^* \| \end{aligned} \qquad (20)$$

Thus it is clear from Eq. (20) that not only the modified PFBS method Eq. (15) and Eq. (16) is convergent, but also it has a linear convergence rate with $\|I-sFA\|$. Thus the smaller value $\|I-sFA\|$ provides faster convergence rate.

*2.3. Fused Analytical and Iterative Reconstruction*

In this section we develop the AIR method in the context of circular cone-beam CT (CBCT), based on the above modified PFBS method with a constant *s*. Specifically, *F* is the FDK operator for the proof-of-concept AIR method for CBCT in this study. That is,

$$x^{n+1/2} = x^n - sF(Ax^n - y), \qquad (21)$$

$$x^{n+1} = \arg\min_x \frac{1}{2} \| x - x^{n+1/2} \|_2^2 + s\lambda \| \nabla x \|_1, \qquad (22)$$

where *F* is the FDK operator [29]. That is,

$$FDK(y) = B \cdot C \cdot Wy, \qquad (23)$$

where *W* is the data pre-weighting operator, *C* the filtering operator (i.e., row-by-row 1D convolutions on weighted projection data with 1D filter), and *B* the backprojection operator. In this work, we also consider the half-fan CBCT scan, for which a data pre-weighting method [30] is used to reduce the artifacts due to large detector shifts.

In the above two-step iterative scheme Eq. (21) and (22), the data fidelity and image regularization are completely decoupled: an FDK-projection step updates the filtered data fidelity term, while a denoising solver updates the sparsity regularization term. During the FDK-projection step, the image is projected to the data domain to form the data residual, and then reconstructed by FDK to a residual image which is then weighted together with previous image iterate to form next image iterate.

On the other hand, since the AIR method via PFBS Eq. (15) and (16) is a general scheme that is not specific to a particular AR method. The method should be generally applicable to other CT image reconstructions other than CBCT, e.g., helical CT, as long as a proper AR method is chosen so that Eq. (19) is satisfied. However, this can be easily done since any AR method is often an inverse or approximate inverse of the X-ray transform.

An advantage of such a choice is that the AIR method Eq. (21) and Eq. (22) has a faster convergence than the IR method Eq. (13) and Eq. (14). This is because *F* is approximately the inverse of *A*, and therefore $\|I-sFA\|$ can be much smaller than $\|I-sA^TA\|$, which guarantees the convergence rate by Eq. (20). In the result section, we will numerically compute $\|I\text{-}sFA\|$ to further verify the faster convergence enjoyed by the AIR method via the modified PFBS.

Next we provide a solution algorithm for solving the proximal operator or the denoising subproblem Eq. (22) with TV regularization using ADMM [24] (also known as split Bregman method [25]). For the convenience of the discussion, we rewrite the denoising problem as

$$x^* = \arg\min_{x} \frac{1}{2} \| x - y \|_2^2 + \lambda \| \nabla x \|_1 . \tag{24}$$

Note that the TV norm Eq. (5) is non-differentiable when TV components are all zeros. One can certainly introduce a small perturbation parameter to make TV differentiable [12], which however may introduce further complication as an approximate TV, such as how to choose the perturbation parameter. Here we solve the exact TV. Although TV is non-differentiable, Eq. (24) is still convex and therefore ADMM can solve it efficiently. The key idea of ADMM for solving Eq. (24) is to decouple L1-norm TV from L2-norm data fidelity, and then the L1-norm minimization Eq. (28) has the analytical solution (i.e., so-called isotropic shrinkage formula) Eq. (34) while the TV transform goes into the L2 data fidelity subproblem Eq. (27).

To start with, we introduce a dummy variable $z$ for the TV transform so that Eq. (20) is transformed to a equivalent constrained optimization problem

$$(x^*, z^*) = \arg\min_{(x,z)} \frac{1}{2} \| x - y \|_2^2 + \lambda \| z \|_1, \nabla x = z . \tag{25}$$

Eq. (25) essentially decouples the TV transform (in variable $x$) from the L1 norm (in variable $z$), and it can be solved by alternately minimizing its following augmented Lagrangian $L$ with respect to $x$, $z$ and $u$,

$$L(x,z,u) = \frac{1}{2} \| x - y \|_2^2 + \lambda \| z \|_1 + \frac{\mu}{2} \| \nabla x - z + u \|_2^2 , \tag{26}$$

where $u$ is a dual variable for the equality constraint in Eq. (25) with an algorithm parameter $\mu$. Then Eq. (25) can be solved via the following alternating ADMM loop

$$x^{m+1} = \arg\min_{x} L(x, z^m, u^m) , \tag{27}$$

$$z^{m+1} = \arg\min_{z} L(x^{m+1}, z, u^m) , \tag{28}$$

$$u^{m+1} = u^m + \nabla x^{m+1} - z^{m+1} . \tag{29}$$

Here Eq. (29) updates the dual variable $u$ to ensure the satisfaction of the equality constraint through iterations [24,46].

Eq. (27) is differentiable and its solution satisfies the following optimal condition

$$(I + \mu \nabla^T \nabla) x^{m+1} = y + \mu \nabla^T (z^m - u^m), \tag{30}$$

which is approximately solved by the one-step conjugate gradient method [46]

$$x^{m+1} = x^m - \alpha^m (Px^m - b^m) \tag{31}$$

with

$$\alpha^m = \frac{(r^m, r^m)}{(r^m, P \cdot r^m)} \text{ and } r^m = Px^m - b^m, \tag{32}$$

when Eq. (30) is abbreviated as $Px^{m+1}=b^m$. Note that $P$ is symmetric and positive definite. Approximate solutions such as Eq. (31) for the L2 subproblem Eq. (27) are commonly used in the iterative scheme Eq. (27)-(29), since they are often as good as exact solutions in terms of the convergence speed [47], for a non-illposed problem such as Eq. (30).

Next we consider the z-subproblem Eq. (28), i.e.,

$$z^{m+1} = \arg\min_z \frac{1}{2} \| z - (\nabla x^{m+1} + u^m) \|_2^2 + \frac{\lambda}{\mu} \| z \|_1. \tag{33}$$

As a reminder, $z$ is a vector field with x, y, and z Cartesian components. Note that the z-subproblem is completely separable in $z$, and each single z-problem can be explicitly solved by the following isotropic shrinkage formula $S$ that can be derived from its optimal condition,

$$z^{m+1} = S_{\lambda/\mu}(\nabla x^{m+1} + u^m). \tag{34}$$

Specifically, for any nonzero vector field $x=(x_1,x_2,x_3)$ with three scalars, $S$ does the following operation

$$S_\lambda(x) = \frac{(x_1, x_2, x_3)}{\sqrt{x_1^2 + x_2^2 + x_3^2}} \cdot \max(\sqrt{x_1^2 + x_2^2 + x_3^2} - \lambda, 0), \tag{35}$$

and $S_\lambda(x)$ is zero for zero vector field.

*2.4. Connection with Filtered Data Fidelity*

Although the AIR method does not have a corresponding optimization problem to solve in the strict sense. It is closely related to the following filtered data fidelity,

$$f(x) = \frac{1}{2} \| F_0^{1/2} (Ax - y) \|_2^2. \tag{36}$$

In Eq. (36), the filtration weighting term $F_0^{1/2}$ is specific to the AR method under consideration.

Take FDK-based AIR Eq. (21)-(23) for CBCT image reconstruction for example. $F_0^{1/2}$ can be defined as

$$F_0^{1/2} = IFT \cdot R^{1/2} \cdot FT \cdot W, \tag{37}$$

where $R^{1/2}$ is the component-wise square root of the 1D filter in the Fourier domain, $FT$ the 1D row-by-row Fourier transform operator, and $IFT$ the corresponding inverse Fourier transform operator. Therefore,

$$\nabla f = A^T F_0 (Ax - y), \qquad (38)$$

where

$$A^T F_0 = A^T W^T (IFT \cdot R \cdot FT) W. \qquad (39)$$

Next, notice that Eq. (39) is equivalent to the FDK operator Eq. (23) in the PFBS scheme Eq. (21), although Eq. (39) is slightly different from Eq. (23). Specifically, $C=IFT \cdot R \cdot FT$; $A^T$ is replaced by $B$; $W^T$ is ignored. The last two modifications can be justified in the sense of unmatched projection/backprojection pairs [10,11]: since the operator $FA$ does not have negative eigenvalues, the replacement of $A^T F_0$ Eq. (39) by $F=FDK$ Eq. (23) is valid in the iterative scheme Eq. (21). In addition, $FA$ with the filtering accelerates the iterative scheme Eq. (21) and Eq. (22) considerably from $A^T A$ or $BA$, since the eigenvalues are now all clustered near unity [10]. This is also intuitively correct since FDK is an approximate AR method and thus $FA$ is close to the unity operator, which coincides with the previous rigorous justification Eq. (20).

Note that the filtered data fidelity Eq. (36) still assumes the X-ray transform as the forward model, i.e., the same $A$ as in Eq. (2). Thus, with the established connection between the AIR method and filtered data fidelity Eq. (36), it is clear that the AIR method iteratively solves the exact X-ray transform, although it utilizes the FDK which is an approximate AR method.

Next we take a close look at the connection between the AIR method and filtered data fidelity for helical CT, although it has been discussed that the AIR method is applicable to helical CT.

To incorporate FDK-type algorithms (e.g., previous works [33-37] and their references) into AIR, we can use the similar filtration weighting Eq. (37) with modified steps during $F$ that are specific to AR under consideration, such as data weighting and rebinning [33-35]. To incorporate the Katsevich algorithm [38-41] into AIR, we can define $F_0^{1/2}$ as follows

$$F_0^{1/2} = C^{1/2} \cdot W_2 \cdot W_1, \qquad (40)$$

where $W_1$ is the data pre-weighting operator with first-order differentiation and re-sampling, $C^{1/2}$ the filtration operator for Hilbert transform with the square root in the Fourier domain, and $W_2$ is the data post-weighting operator. Then the Katsevich inversion operator follows from $A^T F_0$ by ignoring $W_2$ and $W_1^T$. To incorporate the backprojection-filtration algorithm [3] into AIR, we modify Eq. (36) to be the following inner-product formulation of filtered data fidelity

$$f(x) = \frac{1}{2}(x, CA^T W_1 (Ax - y)), \qquad (41)$$

where $W_1$ is the data pre-weighting operator with first-order differentiation, C is 1D filtering along the PI line (i.e., the straight line that connects any two points on the helical trajectory within one turn), and $A^T$ can be replaced by the backprojection operator $B$ along PI lines.

To summarize, for helical CT we again simply need to set $F$ in Eq. (15) to be the corresponding AR operator, as long as Eq. (19) is satisfied. This is also clear based on the filtered data fidelity, which then can be justified by the use of unmatched projection/backprojection pairs [10,11]. In addition, it is straightforward to formulate AIR based on Eq. (37) and (40) for Grangeat's method [42], general inversion formulas based on Riesz potential and other AR methods [43]. Again, note that regardless whether the underlying AR method is approximate or exact, the AIR method solves for the image reconstruction based on the exact X-ray transform model (i.e., the operator $A$), and therefore is at least as accurate as IR.

*2.5. Implementation*

To summarize, the proposed AIR algorithm via PFBS is as follows.

1: Algorithm parameters: $s$, $\lambda$, $\mu$, $N$, and $M$.

2: Initialization: $x^0 = 0$

3: for $n=1$ to $N$ (*main loop via PFBS*)

4:    $x^{n+1/2} = x^n - sF(Ax^n - y)$

5:    $X^0 = x^{n+1/2}$

6:    for $m=1$ to $M$ (*denoising subproblem via ADMM*)

7:       $X^{m+1} = X^m - \alpha^m(PX^m - b^m)$

8:       $z^{m+1} = S_{s\lambda/\mu}(\nabla X^{m+1} + u^m)$

9:       $u^{m+1} = u^m + \nabla X^{m+1} - z^{m+1}$

10:   end for

11:   $x^{n+1} = X^M$

12: end for

13: Return: $x^* = x^N$

Line 4 updates the data fidelity, i.e., in the sense of filtered data fidelity, where $F$ is some AR method, such as FDK for CBCT in this study. Lines 5 to 11 solves the denoising subproblem related to the image regularization. Here *X* is introduced for the illustration purpose, and it is the same as *x* in implementation.

In this work, *A* is discretized from ray-driven method, and *B* is discretized from pixel-driven method, since $A^T$ may cause numerical artifacts for backprojection [31]. In addition, without memory storage need, *A* and *B* are implemented in parallel as on-the-fly operators, for which more efficient algorithms [32] are available with further reduced computational complexity per thread (i.e., O(1) per thread). However, ray-driven and pixel-driven methods are implemented since they already maximize the utility of the state-of-art single-GPU computing architecture.

Regarding a proper choice of *s*, it is related to the Lipschitz constant *L* of the operator *FA*. That is the PFBS algorithm converges for any $s \leq 2/L$ [48]. When *L* is unknown, it can be found through a line search [49]. In this work, we approximate *L* by numerically evaluating $L=4||FAX||_2/||X||_2$ with *X* an identity image, and then set $s=1/L$. Note that the value of *s* affects the algorithm convergence (e.g., large *s* may cause divergence), but it affects little on reconstructed image quality. In addition, various accelerated schemes are available for improved convergence speed such as the Nesterov method [50] and the Barzilai-Borwein step size [51], although a constant *s* is used in this work. On the other hand, one may also use a posterior estimate method to choose *s* based on Eq. (19), in which the power method can be used to compute the maximal eigenvalue of *I-sFA*.

Regarding the parameters for the denoising subproblem, we take sufficiently large *M=100* to ensure the ADMM convergence, since the denoising steps (lines 7 to 9) are computationally negligible comparing with the data-fidelity step (line 4). On the other hand, *μ=1* empirically in this study. In fact, we tested a fairly big range of values, e.g., *μ=0.01, 0.1*, and *10*, and no significant difference was found in algorithm convergence and image reconstruction quality. Note that the choice of *μ* and *M* here should be robust in general since the denoising subproblem is not specific to any particular CT setting, which comes as a result of the PFBS algorithm.

Regarding the choice of *N*, we again take a sufficient large value *N=20*, considering that *FA* is close to unity. On the other hand, we tested the following stopping criterion [52]

$$d = \frac{(\nabla f, \partial g)}{||\nabla f||_2 ||\partial g||_2}, \qquad (42)$$

and found *d* reached the negative value in very few iterations and was no longer decreasing within 10 iterations. However, the robust cut-off point based on this stopping criterion was not easy for us to select, partially due to the fast convergence of the proposed AIR algorithm. And therefore we used a sufficient large number of iterations instead of a stopping criterion.

The model parameter *λ* in Eq. (3) affects the image quality. In fact, it balances data fidelity and image regularization. For example, from our observation of IR or AIR, contrast-to-noise ratio (CNR) and modulated transfer function (MTF), as two popular quantitative tools for evaluating reconstructed image contrast and resolution, both improve as $\lambda$ increases from zero, and then start to compete with each other as $\lambda$ further increases. As a result, it is difficult to choose $\lambda$ without specifying particular image quality metrics of interest. Therefore, we feel the choice of $\lambda$ should be task-based. In this study, we empirically chose $\lambda$ with a balanced consideration of CNR and MTF.

From our experience, the pre-weighting operator $W$ in Eq. (23) is not essential for standard CBCT, although it is still included in FDK implementation for the result section. On the other hand, for half-fan CBCT, the pre-weighting operator $W$ [30] is essential and improves the image quality with reduced half-fan artifacts.

Regarding the selection of the filtering in this work, we use 1D ramp filter with the Hanning window (a popular choice for CBCT) for enhanced image quality and reconstruction speed, while the image noise is suppressed by TV regularization. Note that the use of image regularization in the AIR or IR method is different from simply denoising the FDK image.

The filtering step in FDK is performed in the Fourier domain with oversampling [53]. That is, we oversample 1D ramp filter with the Hanning window in the Fourier domain with oversamping ratio 2, transform pre-weighted projection data $y$ row-wise into Fourier domain, zero-pad on both ends to double its size, multiple with oversampled 1D filter, transform back to image domain, and then take the real part of the centered half.

In terms of the choice of image regularization, other sparsity transforms than TV may improve the image quality further. Particularly, the wavelet tight frame transform $W$ may be of interest for two reasons. First, the algorithm efficiency may be further improved by using $W^T W = I$ and tight frame based iterative soft thresholding algorithm [54]. Second, the reconstructed image quality may be further improved since the tight frame (e.g., wavelet [61] or tensor framelet [22,55]) provides a variety of sparsity representations including TV that can be task-adaptive [56]. Last, the algorithm efficiency of TV-based AIR may be further improved by a fixed point algorithm based on proximal operator of TV [57].

## 3. Materials and Results

The projection data were acquired from Elekta XVI system with a flat-panel *1024×1024* detector of *0.4mm×0.4mm* pixel size. The source-to-detector and source-to-rotation-center distances are *153.6cm* and *100.0cm*. Full-fan projection data were acquired for a Catphan phantom (Fig. 3 and 4) and half-fan projection data were acquired for a pelvis phantom (Fig. 5). The reconstructed images has the resolution *0.5mm* in all directions.

The image reconstruction methods under comparison are: FDK, IR2, IR, and AIR, where IR2 and IR are from the standard IR model Eq. (1) solved by PFBS and ADMM respectively. Note that although PFBS is efficient for AIR, it is not so for IR2, since the eigenvalues of *BA* are no longer clustered near unity. As a result, not only the image quality from IR2 is degraded, but also its convergence is significantly slower than AIR. While the reconstruction parameters of IR2 are chosen in the similar fashion as AIR, the total iteration number is chosen to be *N=500* to ensure the solution convergence. Alternatively, the standard IR model can be solved directly by ADMM [19,20,55,59,60], namely IR in the section, which provides better image quality from IR2, partially due to iterative data fidelity updates through conjugate gradient method instead of one-step update Eq. (13) for dealing with the ill-conditioned linear system.

First we shall verify that AIR has a faster convergence than IR by numerically computing $||I-sFA||$. The power method [58] was utilized to compute the dominant eigenvalue for *I-sFA*. Here *F* was the transpose of *A* derived from the pixen-driven method in IR, and the FDK operator Eq. (23) in AIR. The convergence rate is shown in Fig. 1, in which AIR attains the minimal value 0.72 while IR has the minimal value 0.93. Redundant values (e.g., the convergence rates larger than 1) are plotted in Fig. 1 to ensure the minimal value has been identified with respect to *s* for IR and AIR respectively. Thus it is established that AIR had a faster convergence than IR.

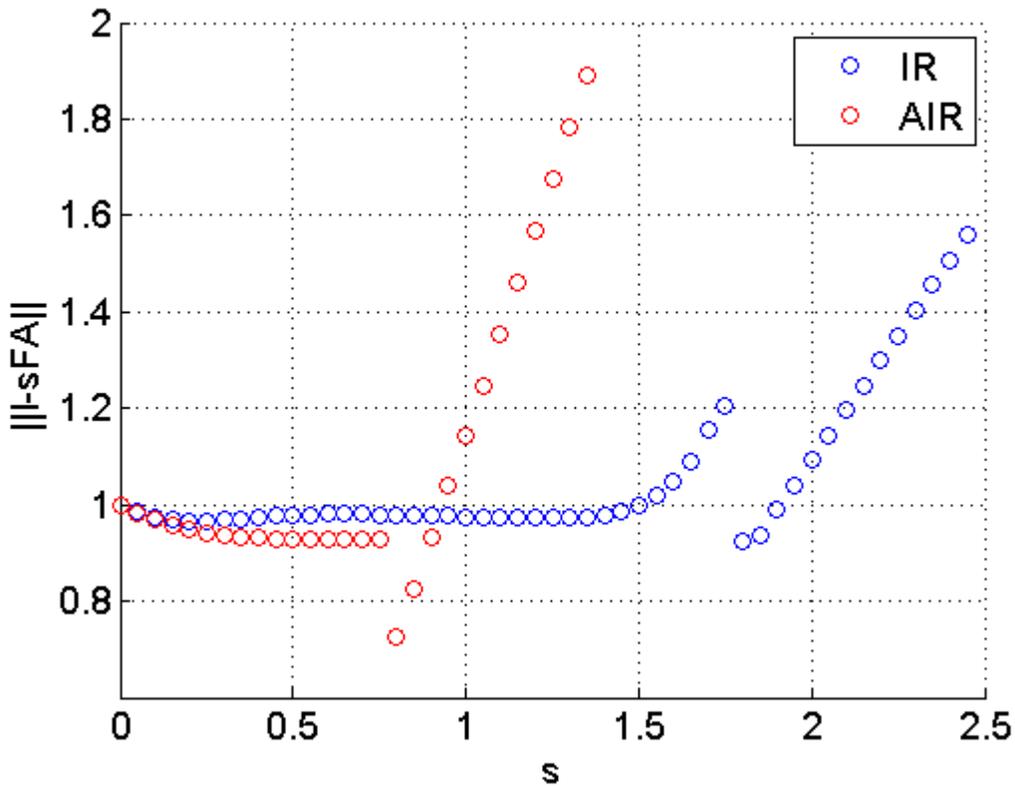

Fig. 1. Convergence rate. AIR attains the minimal value 0.72 while IR has the minimal value 0.93. This verifies that AIR has a faster convergence than IR.

Next we shall compare the image quality. In the following, all the images in Fig. 3-5 are displayed using the same display window. For all these iterative methods, the model parameter $\lambda$ in Eq. (3) that balances data fidelity and image regularization affects the image quality most. However, as discussed in the last section, the optimal choice of $\lambda$ should be task-dependent. In the following, we shall compare the image quality with respect to a sequence of $\lambda$ that are well chosen within a reasonable range. Although there is a trade-off of contrast and resolution with respect to $\lambda$, the image quality is still comparable based on individual trade-off curves (Fig. 2).

To quantify the image quality, we define contrast-to-noise ratio (CNR) and modulated transfer function (MTF) as follows:

$$CNR = \frac{|m_t - m_b|}{\sqrt{\sigma_t^2 + \sigma_b^2}}, \tag{33}$$

$$MTF = \frac{|I_{max} - I_{min}|}{I_{max} + I_{min}}. \tag{34}$$

To quantify the image contrast, we select a contrast slice (Fig. 3) with 7 circular objects. For each object, a target region within a inner circle of the object is selected for computing the mean $m_t$ and standard deviation $\sigma_t$ of pixel values, while a background region within a concentric circle around the object is selected for computing the mean $m_b$ and standard deviation $\sigma_b$ of pixel values. Here selected target and background regions are the same for all the methods. For the ease of visualization and quantitative comparison, all the CNR values are averaged to form a single CNR number in Fig. 2 and Table 1.

To quantify the image resolution, we select a resolution slice (Fig. 4) and compute MTF for 3 consecutive line pairs, i.e., the line pairs in the zoom-in image of Fig. 4 except the smallest line pair. For each line pair, pixel values are selected from the peak region within bright lines and averaged to form $I_{max}$, while pixel values are selected from the valley region between bright lines and averaged to form $I_{min}$. Here selected pixels have fixed locations for all the methods. For the ease of visualization and quantitative comparison, all the MTF values are averaged to form a single MTF number in Fig. 2 and Table 1.

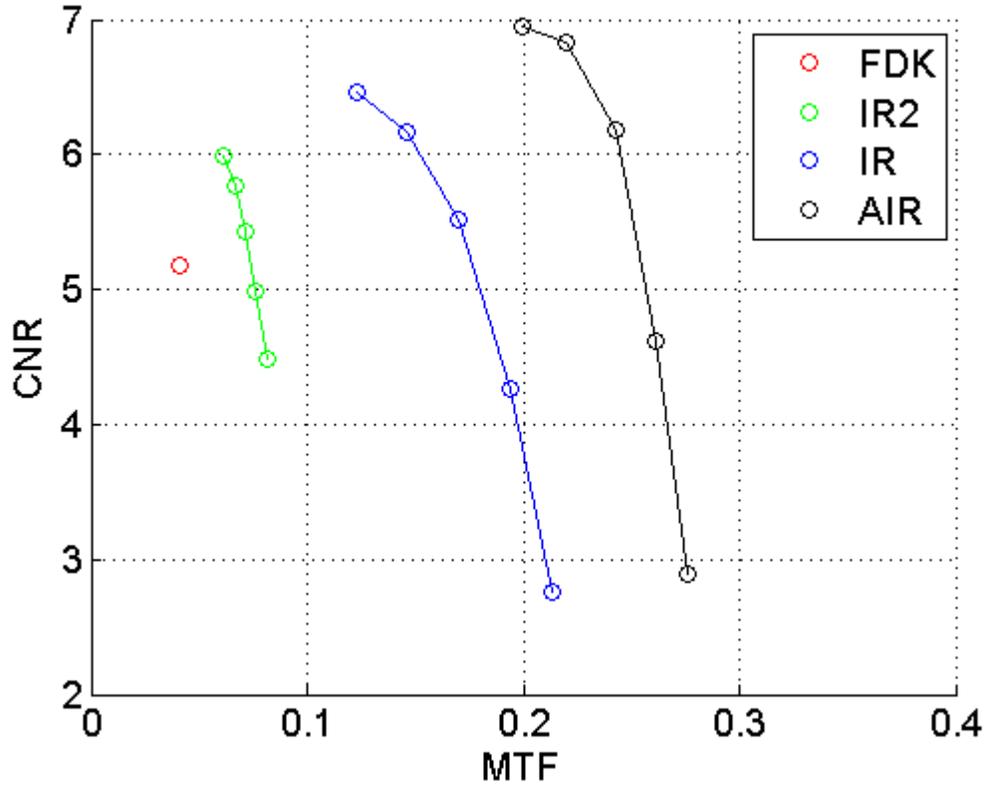

Fig. 2. Quantitative Catphan results. Although there is a trade-off between contrast (CNR) and resolution (MTF) with respect to $\lambda$ for each method, the image quality is still comparable based on individual trade-off curves, i.e., AIR has the best image quality in the sense that it has the best combination of CNR and MTF.

In Fig. 2, the CNR and MTF values are plotted for a equally-spaced sequence of the model parameter $\lambda$, for each of IR2, IR and AIR. Note that the values of $\lambda$ are different for each method. Trade-off curves in Fig. 2 suggest AIR has the best image quality in the sense that it has the best combination of CNR and MTF.

Table 1. Quantitative Catphan results.

|     | FDK   | IR2   | IR    | AIR   |
| --- | ----- | ----- | ----- | ----- |
| CNR | 5.174 | 5.432 | 5.512 | 6.198 |
| MTF | 0.041 | 0.071 | 0.170 | 0.243 |

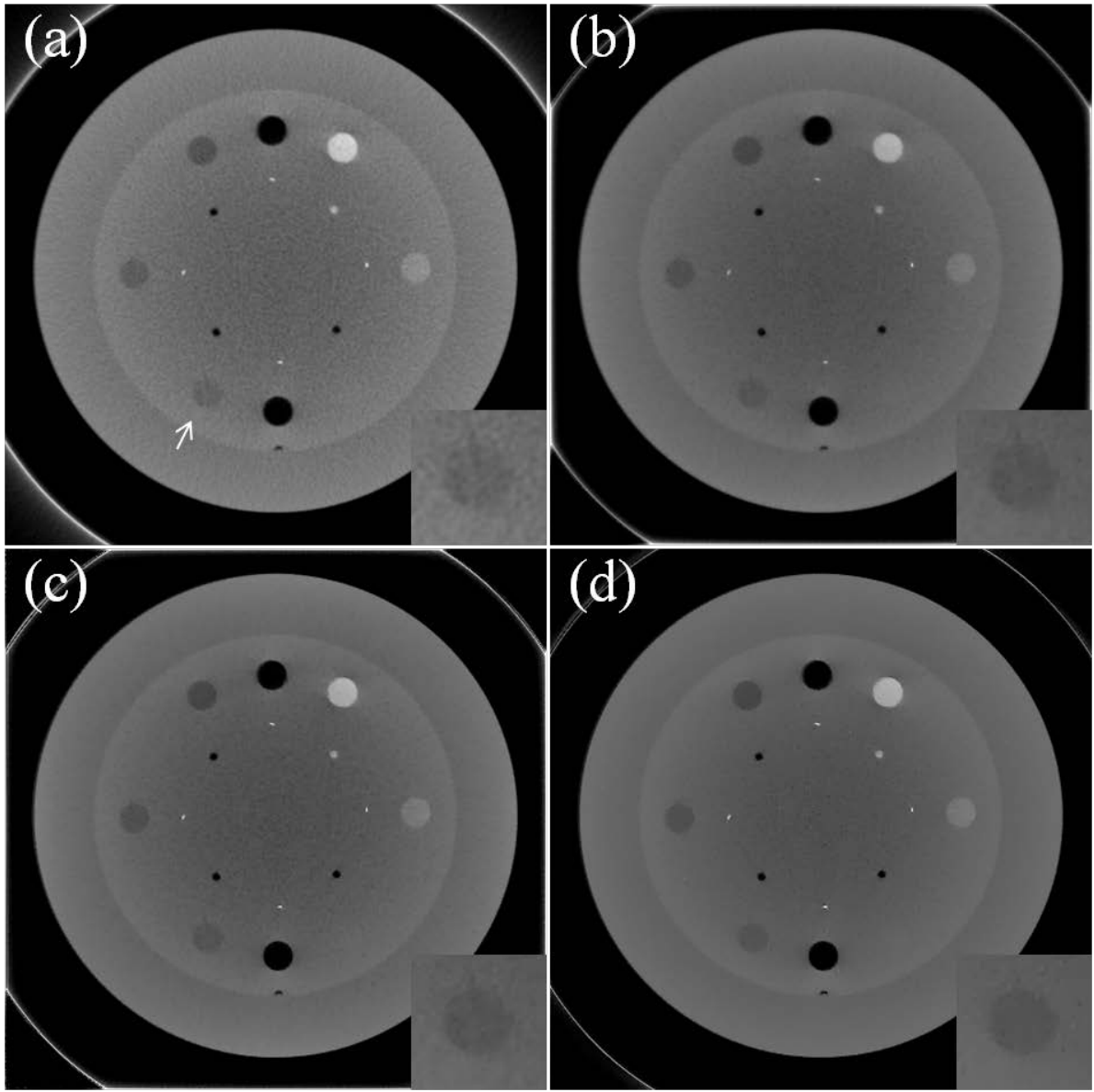

Fig. 3. Catphan results (contrast slice). (a) FDK; (b) IR2; (c) IR; (d) AIR. The corresponding CNR and MTF values are given in Table 1. With the best image resolution (measured by MTF), AIR has comparable image contrast (measured by CNR) with other methods.

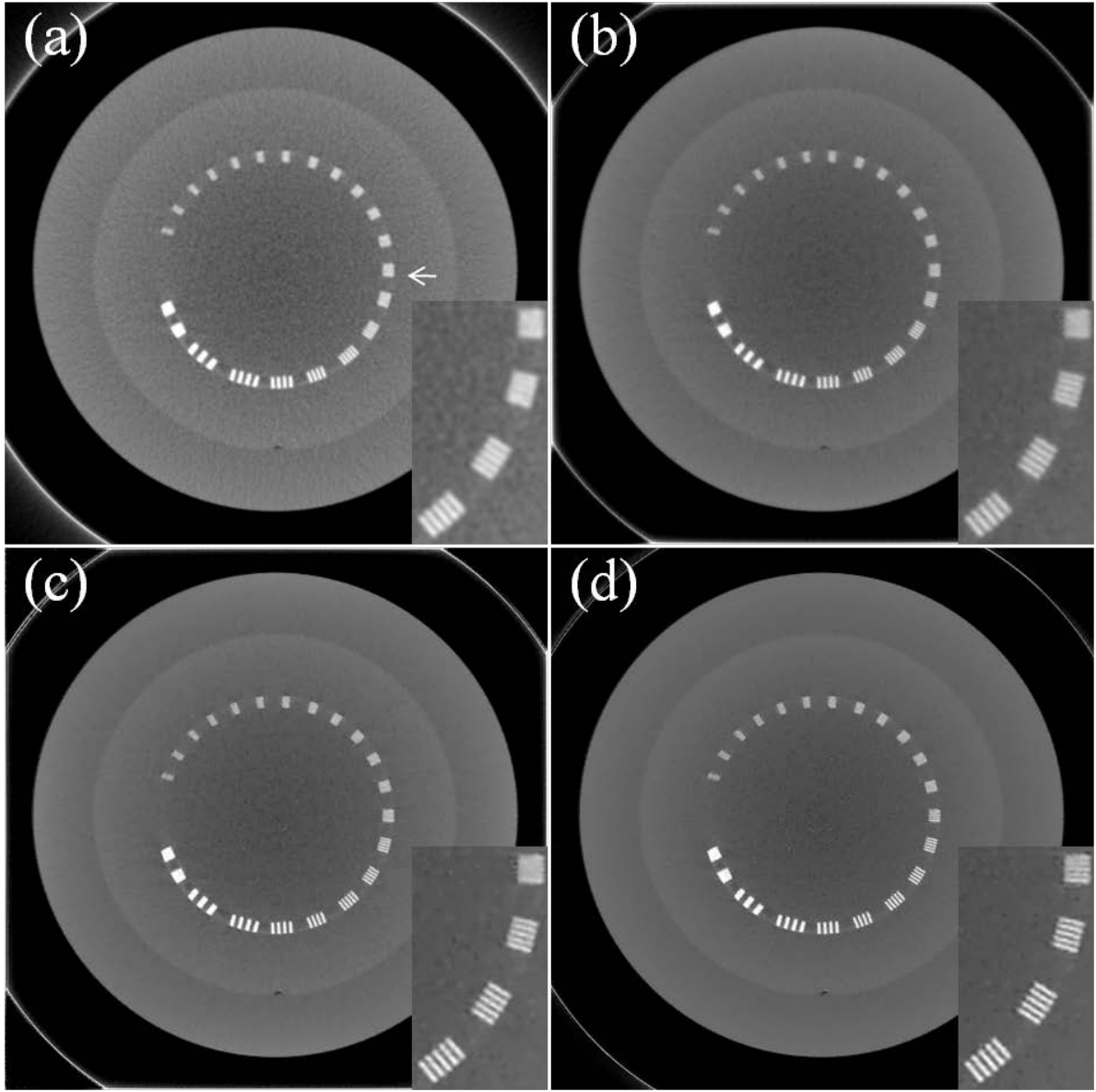

Fig. 4. Catphan results (resolution slice). (a) FDK; (b) IR2; (c) IR; (d) AIR. The corresponding CNR and MTF values are given in Table 1. With comparable image contrast (measured by CNR), AIR has the best image resolution (measured by MTF) with other methods.

Next CNR and MTF from the middle point of IR2, IR and AIR are given in Table 1, together with FDK. Table 1 suggests that AIR has the best image resolution (measured by MTF), under the comparable image contrast (measured by CNR) with other methods. Then the corresponding reconstructed contrast and resolution slices for Table 1 are displayed in Fig. 3 and Fig. 4. By the visual inspection, AIR has improved contrast for a low-contrast object (marked by an arrow in Fig. 3(a)), which is also confirmed from its zoom-in figure; in addition, AIR also has improved resolution from the zoom-in line pairs.

The pelvis results with half-fan projection data are plotted in Fig. 5. Due to large detector shifts, the half-fan artifact severely degraded the reconstructed image quality by FDK (Fig. 5(a) and 5(b)). On the other hand, although the half-fan artifact was alleviated by IR, it generated artificial high-contrast rings following the trajectory of detector edges during half-fan acquisition (Fig. 5(e)). Moreover, IR was prone to the axial artifact that appeared at boundary axial slices (Fig. 5(f)).

By modifying FDK using a data pre-weighting method [30], the half-fan artifact was reduced (Fig. 5(c) and 5(d)), which however still generated an artificial shade in the center that was redundantly covered during the half-fan scan.

In contrast, by incorporating this modified FDK with data-pre-weighting, AIR significantly improved the reconstructed image quality (Fig. 5(g) and 5(h)). By the visual inspection, AIR again has the best combination of contrast and resolution. In particular, not only the axial artifact, i.e., boundary artifact along the axial direction in the sagittal slice (Fig. 5(f)) in IR was completely eliminated in AIR, but also the half-fan artifact, i.e., the dark region in modified FDK (Fig. 5(c) and 5(d)) or the bright circle in IR (Fig. 5(e) and 5(f)) was almost eliminated in AIR (Fig. 5(g) and 5(h)). As a result of reduced half-fan artifact, the image features that are blurred in FDK or IR are now visible in AIR, e.g., these zoom-in ROI images in Fig 5.

Note that the half-fan artifact can also be reduced by incorporating a data-weighting function into IR [62]. In this case, the backprojection operator is no longer the transpose of the projection operator, for which the convergence is exactly what has been considered in this paper, i.e., see the modified PFBS algorithm in Section 2.2. On the other hand, the axial artifact from IR can be reduced for certain ROI at sagittal boundary by enlarging the reconstruction volume, which pushes the axial artifact further away axially at the expense of increased computational cost.

To summarize, the proposed AIR improved reconstructed image quality in terms of both resolution and contrast from AR or IR, e.g., the Catphan results (Fig. 2-4). Moreover, AIR has the flexibility to naturally incorporate the data weighting into the AR operator for reduced imaging artifact (Fig. 5), e.g., half-fan artifact and axial artifact.

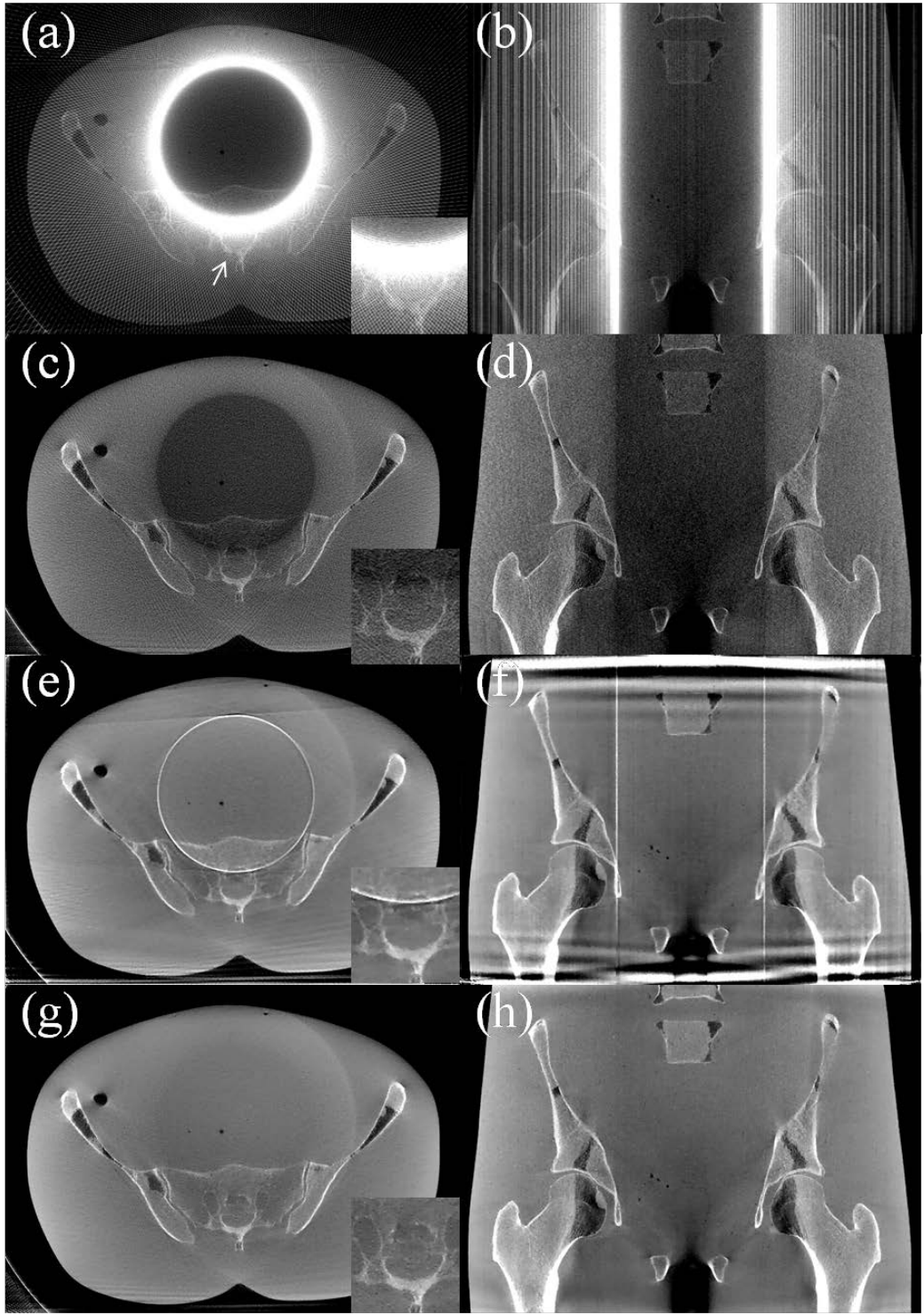

Fig. 5. Pelvis results. FDK: (a) and (b); modified FDK: (c) and (d); IR: (e) and (f); AIR: (g) and (h). Transverse slice: (a), (c), (e) and (g); Coronal slice: (b), (d), (f) and (h). By incorporating the modified FDK with data-pre-weighting, AIR improved the reconstructed image quality with significantly reduced half-fan artifact and axial artifact.

## 3. Conclusion

AIR is proposed to incorporate AR into IR, with an efficient image reconstruction algorithm based on PFBS. The convergence of PFBS-based AIR has been established with superior convergence rate, and its connection with filtration-weighted data fidelity has been discussed. Based on PFBS, AIR provides a two-step iterative scheme with decoupled treatment of data fidelity and image regularization through a AR-projection update and a denoising solver respectively. As a result, PFBS based AIR has fast convergence since the eigenvalues of the AR-projection operator are close to the unity. In addition, the incorporation of various AR methods is discussed for AIR. The CBCT results suggest that AIR synergizes AR and IR with improved image quality and reduced axial and half-fan artifacts.

## Acknowledgment

The author is grateful to the reviewers for their valuable comments. I thank Dr. Yang-Kyun Park from Massachusetts General Hospital for projection data and inspiring discussions, Xue Zhang from Shanghai Jiao Tong University for convergence analysis, and Jiulong Liu from Shanghai Jiao Tong University for numerical computation of convergence rate.